\documentclass[fleqn,usenatbib]{mnras}

\usepackage{newtxtext,newtxmath}

\usepackage[T1]{fontenc}

\DeclareRobustCommand{\VAN}[3]{#2}
\let\VANthebibliography\thebibliography
\def\thebibliography{\DeclareRobustCommand{\VAN}[3]{##3}\VANthebibliography}

\usepackage{graphicx}	
\usepackage{amsmath}	
\usepackage{verbatim}
\usepackage{gensymb}    

\title[Evaporation before disruption]{Evaporation before disruption: comparing timescales for Jovian planets in star-forming regions}

\author[E. C. Daffern-Powell \& R. J. Parker]{
Emma C. Daffern-Powell and Richard J. Parker\thanks{E-mail: R.Parker@sheffield.ac.uk}\thanks{Royal Society Dorothy Hodgkin Fellow}
\\
Department of Physics and Astronomy, The University of Sheffield, Hicks Building, Hounsfield Road, Sheffield S3 7RH, UK}

\pubyear{2022}

\begin{document}
\label{firstpage}
\pagerange{\pageref{firstpage}--\pageref{lastpage}}
\maketitle

\begin{abstract}
Simulations show that the orbits of planets are readily disrupted in dense star-forming regions; planets can be exchanged between stars, become free-floating and then be captured by other stars. However, dense star-forming regions also tend to be populous, containing massive stars that emit photoionising radiation, which can evaporate the gas in protoplanetary discs. We analyse $N$-body simulations of star-forming regions containing Jovian-mass planets and determine the times when their orbits are altered, when they become free-floating, and when they are stolen or captured. Simultaneously, we perform calculations of the evolution of protoplanetary discs when exposed to FUV radiation fields from massive stars in the same star-forming regions. In almost half (44\,per cent) of the planetary systems that are disrupted -- either altered, captured, stolen or become free-floating, we find that the radius of the protoplanetary disc evolves inwards, or the gas in the disc is completely evaporated, before the planets' orbits are disrupted. This implies that planets that are disrupted in dense, populous star-forming regions are more likely to be super Earths or mini Neptunes, as Jovian mass planets would not be able to form due to mass loss from photoevaporation. Furthermore, the recent discoveries of distant Jovian mass planets around tightly-packed terrestrial planets argue against their formation in populous star-forming regions, as photoevaporation would preclude gas giant planet formation at distances of more than a few au.  
\end{abstract}





\section{Introduction}

Most stars form in groups \citep{Lada03,Bressert10} where the stellar density siginficantly exceeds that of the Galactic field by several orders of magnitude \citep{Korchagin03}. Present-day densities in star-forming regions span the range ($\sim 10 - 10^3$\,M$_\odot$\,pc$^{-3}$), but the \emph{initial} densities may be higher still \citep{Marks12,Parker22a}. In addition, most star-forming regions form with spatial and kinematic substructure \citep{Cartwright04,Sanchez09}, which increases the chances of interactions and encounters in the early stages of a star's life.

The formation of planetary systems occurs contemporaneously with the star formation process, with dust and gas-rich protoplanetary discs \citep{Haisch01,Richert18} ubiquitous around young ($<10$\,Myr) stars, and observations indicating that the discs contain substructures that may be signatures of planetary systems within them \citep{Brogan15,Andrews18,Alves20,SeguraCox20}.

The relatively high stellar densities, combined with non-equilibrium initial conditions in the spatial and kinematic substructure, means that planetary systems can be disrupted in their birth environments. At the highest stellar densities ($\geq 10^4$M\,$_\odot$\,pc$^{-3}$), direct truncation of protoplanetary discs can occur \citep{Vincke16,Winter18b}, and at more modest stellar densities ($\geq 100$M\,$_\odot$\,pc$^{-3}$) direct disruption of planetary orbits occurs \citep{Smith01,Adams06,Parker12a,DaffernPowell22}.

However, if massive stars ($>5$\,M$_\odot$) are present in a star-forming region, the Far Ultraviolet (FUV) and Extreme Ultraviolet (EUV) radiation emitted by these stars can photovaporate the gas content of protoplanetary discs \citep{Scally01,Adams04,Fatuzzo08,ConchaRamirez19,Nicholson19a,Parker21a}. Whilst the dust content is largely unaffected by this photoevaporation \citep{Haworth18b}, the planetary systems that are able to form in star-forming regions containing massive stars may be devoid of gas giant planets like Jupiter and Saturn.

Much of the literature on planetary disruption in dense stellar environments focusses on the effects of encounters on Jupiter-mass planets \citep[e.g.][]{Parker12a}, but star-forming regions with densities high enough to alter or disrupt the orbits of gas giants would also generate high FUV and EUV fluxes from the massive stars. Furthermore, photoevaporation is an extremely fast ($<1$\,Myr) process in regions where the stellar densities would be high enough to alter the orbits of Jupiter- and Saturn-mass planets \citep{Parker21a}. Whilst it is not possible to self-consistently model the full planet formation process and track dynamical encounters in star-forming regions, in principle it is possible to compare the timescale for diruption due to dynamical encounters to the timescale for disc destruction due to photoevaporation. 

In this paper, we determine the times at which gas giant planets have their orbits altered, or become free-floating, in simulated star-forming regions, and compare this to the timescale for photoevaporation of the disc from which the gas giants form. We describe our simulations, and disc photoevaporation analysis, in Section~\ref{sec:Methods}, we present our results in Section~\ref{sec:ResultsDiscussion} and we draw conclusions in Section~\ref{sec:conclusions}.

\section{Methods}
\label{sec:Methods}

We couple $N$-body simulations of the dynamical evolution of star-forming regions that contain a population of Jupiter-mass planets with a post-processing analysis where we follow the evolution of the protoplanetary discs in the presence of photoionising radiation fields.

\subsection{$N$-body simulations}

We use two sets of the $N$-body simulations described in \citet{DaffernPowell22}. We focus on the most dense simulations so that we can determine the maximum impact of both dynamical encounters that would alter/disrupt planetary orbits, and the highest radiation fields that would disrupt/destroy the protoplanetary discs. However, we also analyse a set of lower density simulations. Both sets of simulations contain $N_\star = 1000$ stars, drawn from a \citet{Maschberger13} IMF with a probability distribution of the form
\begin{equation}
p(m) \propto \left(\frac{m}{\mu}\right)^{-\alpha}\left(1 + \left(\frac{m}{\mu}\right)^{1 - \alpha}\right)^{-\beta}.
\label{maschberger_imf}
\end{equation}
Here, $\mu = 0.2$\,M$_\odot$ is the average stellar mass, $\alpha = 2.3$ is the \citet{Salpeter55} power-law exponent for higher mass stars, and $\beta = 1.4$ describes the slope of the IMF for low-mass objects \citep*[which also deviates from the log-normal form;][]{Bastian10}. We randomly sample this distribution in the mass range 0.1 -- 50\,M$_\odot$, such that brown dwarfs are not included in the simulations. This distribution is sampled stochastically, so different realisations of the same simulation contain different numbers of massive stars, but we typically obtain 5 -- 20 stars with masses $>5$\,M$_\odot$ that will produce photoionising radiation.

For simplicity (and to reduce computational expense) we do not include primordial stellar binaries, although these are ubiquitous in star-forming regions \citep{Duchene13b}. Half of the stars are randomly assigned a 1\,M$_{\rm Jup}$ planet with semimajor axis $a_p = 5$\,au and zero eccentricity, i.e.\,\,a Jupiter-like orbit. Whilst the occurrence rate of gas giant planets in extrasolar systems may not be has high as 50\,per cent, we simply aim to test how often the gas content of the protoplanetary disc would be destroyed before the gas giant orbit is disrupted (if at all), and the number of gas giants does not affecte the overall dynamical evolution of the star-forming region.

  We do not allow stars with mass $\geq$3\,M$_\odot$ to host planets. Recent observational work suggests that massive stars can host planets \citep{Janson21b}, but their formation mechanism is unclear and could be dynamical, rather than the planets forming in discs around the massive stars \citep{Parker22b}.

We also note that we are assuming our 1\,M$_{\rm Jup}$ planets are able to form quickly, before the start of our $N$-body simulations. This is a strong assumption, although we note that the planets could accrete from their protoplanetary discs during the evolution of the star-forming regions. This is currently beyond the technical capability of our simulations \citep[see][for preliminary research in this area]{Rosotti14}.

The stars (and their planetary systems) are distributed within a box-fractal distribution \citep{Goodwin04a,DaffernPowell20} to mimic the spatial and kinematic substructure observed in many star-forming regions. We adopt a fractal dimension $D=1.6$, which is the higest degree of substructure possible in three dimensions. The velocities are set such that nearby stars have similar velocities (i.e\,\,a small local velocity dispersion), whereas distant stars can have very different velocities, similar to the observed \citet{Larson81} laws. Adopting a high amount of substructure slightly reduces the potency of any photoevaporation  compared to a smoother distribution \citep{Parker21a}.

We set the radius of the fractals to be either $r_F = 1$\,pc, resulting in a median local stellar density in the fractals of $\tilde{\rho} \sim 10^4$M$_\odot$\,pc$^{-3}$, or $r_F = 5$\,pc, which produces a more modest median local stellar density of $\tilde{\rho} \sim 10^2$M$_\odot$\,pc$^{-3}$.

  Observations \citep[e.g.][]{Parker17a,Sacco17} suggest many local star-forming regions have densities towards the lower values, but more distant, populous star-forming regions may have much higher densities \citep{Schoettler22}. Planetary orbits are disrupted above densities of $\tilde{\rho} \sim 10^2$M$_\odot$\,pc$^{-3}$ \citep{Bonnell01,Adams06,Parker12a}, and photoevaporation can destroy discs at even lower densities \citep{Parker21a}, so we will determine whether the choice of initial density affects our results.

We scale the velocities of the stars such that the global virial ratio is  $\alpha_{\rm vir} = T/|\Omega|$, where $T$ and $|\Omega|$ are the total kinetic and potential energies, respectively. The velocities of young stars are often observed to be subvirial along filaments, so we adopt a subvirial ratio ($\alpha_{\rm vir} = 0.3$) in all of our simulations.

To assess the statistical significance, we run 20 realisations of the same simulation, identical apart from the random number seed used to initialise the initial mass, velocity and position distributions. The simulations are evolved for 10\,Myr using the \texttt{kira} integrator within the \texttt{Starlab} environment \citep{Zwart01}. We do not include stellar evolution in the simulations. Data are outputed as snapshots at intervals of 0.01\,Myr.  

\subsection{Photoevaporation and disc evolution}

We perform a post-processing analysis to determine the effects of photoevaporation on our planetary systems. In other words, the planetary systems are allowed to dynamically evolve independently of the discs, and we then determine how many of the planetary systems would have undergone significant photoevaporation before the planets are then dynamically disrupted.

We achieve this by calculating the FUV flux incident on each low-mass star,
\begin{equation}
 F_{\rm FUV} = \frac{L_{\rm FUV}}{4\pi d^2},
 \end{equation}
where $d$ is the distance from each low-mass star to each star more massive than 5\,M$_\odot$. The simulations all contain more than one massive star, so we sum these fluxes to obtain the FUV radiation field for each disc-bearing star. This FUV radiation field is then scaled to the \citet{Habing68}  unit, $G_0 = 1.8 \times 10^{-3}$\,erg\,s$^{-1}$\,cm$^{-2}$, which is the background FUV flux in the interstellar medium. The FUV luminosities are taken from \citet{Armitage00}.

For each planet-hosting star (masses between $0.1 < M_\star/{\rm M_\odot} < 3$\,M$_\odot$), we assign it a disc of mass
 \begin{equation}
   M_{\rm disc} = 0.1\,M_\star,
 \end{equation}
 and a radius $r_{\rm disc} = 50$\,au. This is so that the disc radius is comfortably larger than the initial semimajor axes of the planets (5\,au). 

 To calculate the mass loss due to FUV radiation, we use the \texttt{FRIED} grid of models from \citet{Haworth18b}, which uses the stellar mass, $M_\star$, radiation field $G_0$, disc mass $M_{\rm disc}$ and disc radius $r_{\rm disc}$ as an input, and produces a mass-loss rate, $\dot{M}_{\rm FUV}$ as an output. As the \texttt{FRIED} grid produces discrete values, we perform a linear interpolation over disc mass and mass-loss. 

In addition to calculating the mass-loss due to FUV radiation, we also calculate the (usually much smaller) mass loss due to EUV radiation. To calculate the mass loss due to EUV radiation, we adopt the following prescription from \citet{Johnstone98}:
\begin{equation}
\dot{M}_{\rm EUV} \simeq 8 \times 10^{-12} r^{3/2}_{\rm disc}\sqrt{\frac{\Phi_i}{d^2}}\,\,{\rm M_\odot \,yr}^{-1}.
\label{euv_equation}
\end{equation}
Here, $\Phi_i$ is the  ionizing EUV photon luminosity from each massive star in units of $10^{49}$\,s$^{-1}$ and is dependent on the stellar mass according to the observations of \citet{Vacca96} and \citet{Sternberg03}. For example, a 41\,M$_\odot$ star has $\Phi = 10^{49}$\,s$^{-1}$ and a 23\,M$_\odot$ 
star has $\Phi = 10^{48}$\,s$^{-1}$. The disc radius $r_{\rm disc}$ is expressed in units of au and the distance to the massive star $d$ is in pc.

  We subtract mass from the discs according to the FUV-induced mass-loss rate in the \texttt{FRIED} grid and the EUV-induced mass-loss rate from Equation~\ref{euv_equation}. Models of mass loss in discs usually assume the mass is removed from the edge of the disc (where the surface density is lowest) and
 we would expect the radius of the disc to decrease in this scenario. We employ a very simple way of reducing the radius by assuming the surface density of the disc at 1\,au, $\Sigma_{\rm 1\,au}$, from the host star remains constant during mass-loss \citep[see also][]{Haworth18b,Haworth19}. If
  \begin{equation}
\Sigma_{\rm 1\,au} = \frac{M_{\rm disc}}{2\pi r_{\rm disc} [{\rm 1\,au}]},
\label{rescale_disc}
  \end{equation}
  where $M_{\rm disc}$ is the disc mass, and $r_{\rm disc}$ is the radius of the disc, then if the surface density at 1\,au remains constant, a reduction in mass due to photoevaporation will result in the disc radius decreasing by a factor equal to the disc mass decrease. 

The decrease in disc radius due to photoevaporation will be countered to some degree by expansion due to the internal viscous evolution of the disc. We implement a very simple prescription for the outward evolution of the disc radius due to  viscosity following the procedure in \citet{ConchaRamirez19a}. 

First, we define a temperature profile for the disc, according to
\begin{equation}
T(R) = T_{\rm 1\,au} R^{-q},
\end{equation}
where $R$ is the distance from the host star, $T_{\rm 1\,au}$ is the temperature at 1\,au from the host star and is derived from the stellar luminosity. For a 1\,M$_\odot$ star, we derive $T_{\rm 1\,au} = 393$\,K, although $T_{\rm 1\,au} = 300$\,K is more commonly adopted. We assume a main sequence mass-luminosity relation, and use data from \citet{Cox00}. We adopt $q = 0.5$ \citep{Hartmann98}.  

 In the model of \citet{Hartmann98}, the characteristic initial radius, $R_c(0)$ is defined by 
 \begin{equation}
 R_c(0) = R'\left(\frac{M_\star}{{\rm M_\odot}} \right)^{0.5},
 \end{equation} 
 where $R' = 30$\,au. At some time $t$, the characteristic radius $R_c(t)$ at that time is given by \citep{LyndenBell74}
  \begin{equation}
R_c(t) = \left( 1 + \frac{t}{t_\nu}\right)^{\frac{1}{2 - \gamma}} R_c(0), 
 \end{equation}
 where the viscosity exponent $\gamma$ is unity \citep{Andrews10}. $t_\nu$ is the viscous timescale, and is given by
  \begin{equation}
t_\nu =   \frac{\mu_{\rm mol}m_pR_c(0)^{0.5 + q}\sqrt{GM_\star}}{3\alpha(2 - \gamma)^2k_BTR^q},
 \end{equation}
 where $\mu_{\rm mol}$ is the mean molecular weight of the material in the disc (we adopt $\mu_{\rm mol} = 2$), $m_p$ is the proton mass, $G$ is the gravitational constant, $M_\star$ is the mass of the star, $k_B$ is the Boltzmann constant and $T$ and $R$ are the temperature and distance from the host star, as described above.  $\alpha$ is the turbulent mixing strength \citep{Shakura73} and based on observations of T~Tauri stars, \citet{Hartmann98} adopt $\alpha =  10^{-2}$.
 
 We set $r_{\rm disc} = R$ to be the radius of the disc, and following mass-loss due to photoevaporation and the subsequent inward movement of the disc radius according to Equation~\ref{rescale_disc}, we calculate the change in characteristic radius ($R_c(t_n)/R_c(t_{n-1})$) and scale the disc radius $r_{\rm disc}$ accordingly:
 \begin{equation}
 r_{\rm disc}(t_n) = r_{\rm disc}(t_{n - 1})\frac{R_c(t_n)}{R_c(t_{n-1})}.
 \end{equation} 

 Both the inward evolution of the disc radius due to mass-loss, and the outward viscous evolution occur on much shorter timescales than the gravitational interactions between stars in the star-forming regions. We therefore adopt a timestep of $10^{-3}$\,Myr for the disc evolution calculations \citep{Parker21a}.

\section{Results}
\label{sec:ResultsDiscussion}

We essentially perform two analyses on the data; first, we determine whether a planet has had its orbit altered (but still orbiting its parent star), or if a planet has been captured (i.e.\,\,has been free-floating for at least 0.01Myr before being (re)captured) or stolen (directly exchanged between two stars without ever being free-floating), or whether a planet has been liberated and is free-floating in the snapshot.

If the planet is bound to a star, we record its semimajor axis, or if it is free-floating, we record the semimajor axis at the point at which it became free-floating. 

Simultaneously, for each planet hosting star (or former host star if the planet is now free-floating), we determine the evolution of the disc radius if subject to mass-loss caused by photoevaporation. In the initially high density simulations ($\tilde{\rho} = 10^4$\,M$_\odot$\,pc$^{-3}$) the initial median FUV fields in our simulations are of order $10^4 G_0$, decreasing to several 100$G_0$ \citep{Parker21a}. In the initially moderate density simulations ($\tilde{\rho} = 10^2$\,M$_\odot$\,pc$^{-3}$) the initial median FUV fields are of order $10^3 G_0$, but remain relatively constant during the simulation \citep{Parker21a}. 

\subsection{High initial stellar density ($\tilde{\rho} = 10^4$\,M$_\odot$\,pc$^{-3}$)}

We first present our results for simulations which have high initial stellar density  ($\tilde{\rho} = 10^4$\,M$_\odot$\,pc$^{-3}$). In these simulations, the substructure is erased within the first 0.1\,Myr, and the stars fall into the potential well of the star-forming region, forming a bound cluster within 1\,Myr \citep{Allison10,Parker14b}.

\begin{figure}
    \centering
    \includegraphics[width=\columnwidth]{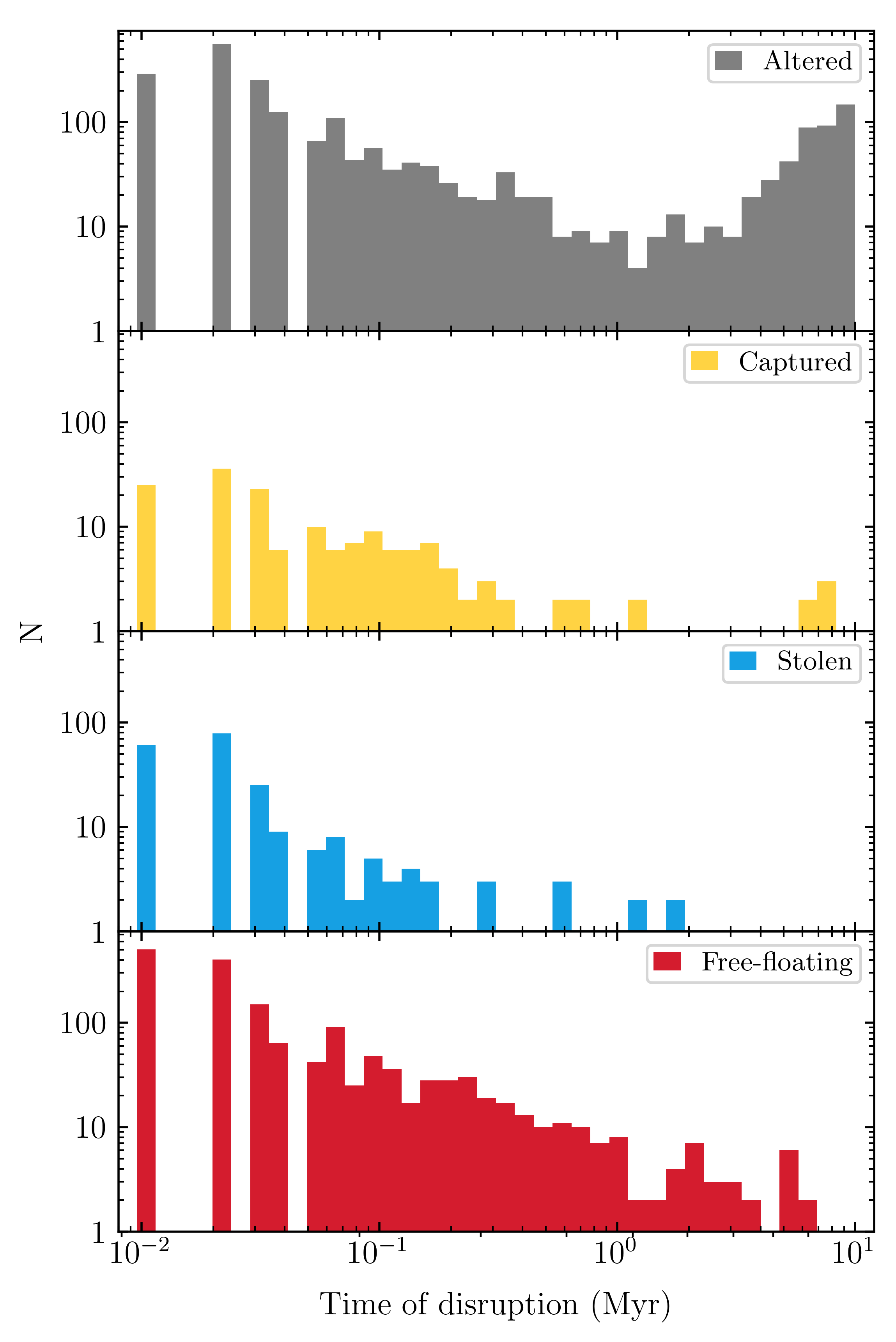}
    \caption{Histograms of the times at which planets are altered, captured, stolen or liberated in our simulations with high initial stellar densities ($\tilde{\rho} = 10^4$\,M$_\odot$\,pc$^{-3}$). The top panel shows planets that remain bound to their host star, but whose eccentricity has changed by more than 0.1, and/or semimajor axis has changed by more than 10\,per cent of the original value. The second panel shows planets that were captured, the third panel shows planets that were stolen by another star and the bottom panel shows planets that have become free-floating.}
    \label{fig:timescales}
\end{figure}

In Fig.~\ref{fig:timescales} we show histograms of the time at which the planets undergo disruption, and  split the figure into panels depending on the mode of disruption. The top panel shows planets that remain bound to their parent stars, but where the eccentricity has changed by more than $\Delta e > 0.1$, and/or the semimajor axis has changed by $\Delta a_p \pm 0.1a_p$. The second panel shows the time at which a planet is (re)captured around a star, having been free-floating for some time ($\geq 0.01$\,Myr) before capture. The third panel shows the times when planets are stolen by another  star \citep[defined as a direct exchange interaction,][]{DaffernPowell22}. The bottom panel shows the times when planets become free-floating (and remain free-floating, i.e.\,\,they are not subsequently captured by another star). The gaps in the histogram at times less than 0.1\,Myr are due to our choice to bin the data in equal logspace, as the data are outputted every 0.01\,Myr, and are not physical gaps in the disruption of planets.  

This plot clearly shows that the majority of planetary disruption occurs early on in the simulation, i.e.\,\,within the first 0.1\,Myr. This is unsurprising, as the first process that occurs in these simulations is that the substructure undergoes violent relaxation \citep{Allison10,Parker14b}, leading to a heightened rate of encounters \citep{DaffernPowell22}.

\begin{figure}
    \centering
    \includegraphics[width=\columnwidth]{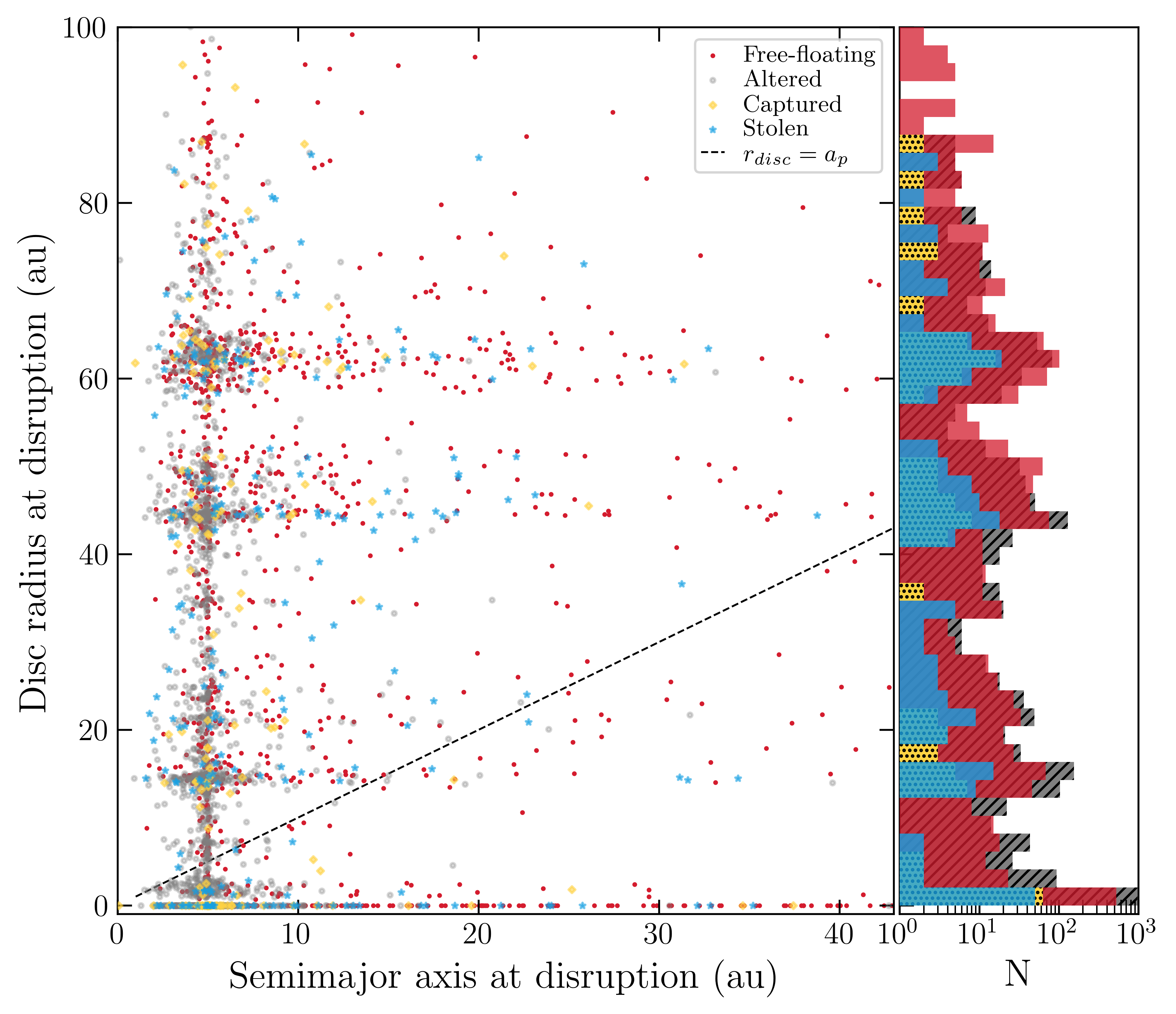}
    \caption{The radius of the hypothetical disc versus the semimajor axis of the planet when its orbit is altered, or when a planet is captured or stolen, or when a planet becomes free-floating, in high density ($\tilde{\rho} = 10^4$\,M$_\odot$\,pc$^{-3}$) simulations. The disc radius is the outer radius of the disc, and the semimajor axis is the instant the planet undergoes an interaction. The dashed line shows $r_{\rm disc} = a_p$; where $a_p > r_{\rm disc}$, the gas in the disc has already been photoevaporated before the planet has undergone a significant dynamical encounter. The histogram indicates the numbers of planets in each category with a disc radius at the instant of disruption. }
    \label{fig:sma_rad}
\end{figure}

However, the majority of mass-loss due to photoevaporation also occurs within the first 0.1\,Myr, and so we might ask whether a disrupted Jupiter-mass planet would have been able to form in the first instance. In Fig.~\ref{fig:sma_rad} we show the disc radius at the instant of planetary disruption as a function of the semimajor axis of the planet at the time it was disrupted. For the captured planets (yellow points) this represents the semimajor axis of the planet immediately after capture, and for the stolen planets (blue points) this is the semimajor axis around their original star immediately before theft by the intruding star. For the free-floating planets (red points), it is the semimajor axis in the instant immediately before being liberated.

To aid interpretation of this plot, we also show the point at which the disc radius is equal to the semimajor axis. Any point to the right of this line represents a planet on a semimajor axis larger than the remaining disc radius, including systems whose discs have been completely destroyed. Of the 10\,000 planets across twenty simulations, 4257 are disrupted -- either altered (2253), stolen (220), captured (171) or become free-floating (1613). Of these 4257 planets, 1871 (44\,per cent) have a semimajor axis that \emph{exceeds} the disc radius at the instant of disruption, and 1441 (34\,per cent of the disrupted systems) have a disc radius of zero.

In Fig.~\ref{fig:sma_rad}, there are noticeable groupings of points, which are an artefact of the discretization of the \texttt{FRIED} grid. When the discs initially lose mass due to photoevaporation, the disc radii decrease, and the next time we access the \texttt{FRIED} grid, the surface density at the edge of the disc has increased because the disc has the same mass, but now a smaller radius. The higher surface density in turn reduces the mass lost due to photoevaporation in the next timestep, and so the disc survivies for longer at this new radius.

  The fraction of systems with semimajor axis larger than the disc radius varies between individual simulations, likely due to the stochastic sampling of the stellar initial mass function (IMF), which leads to different numbers, and individual masses, of the massive stars. This in turn leads to higher or lower radiation fields depending on the number of massive stars.

In our simulations, the two extrema are a simulation where the five most massive stars are 18, 15, 13, 13, \& 11\,M$_\odot$, compared to a simulation with stars of mass 44, 44, 24, 21 \& 18\,M$_\odot$. In the former simulation, the fraction of systems with a semimajor axis greater than the disc radius is 34\,per cent, whereas in the latter it is 42\,per cent. Note that these fractions are both lower than the total fraction across twenty simulations (44\,per cent), so there may not be a straightforward mapping the the numbers of (and massses of) massive stars to the amount of photoevaporation \citep[massive stars can be ejected from these simulations,][which would reduce the FUV field therein]{Schoettler19}.

\begin{figure}
    \centering
    \includegraphics[width=\columnwidth]{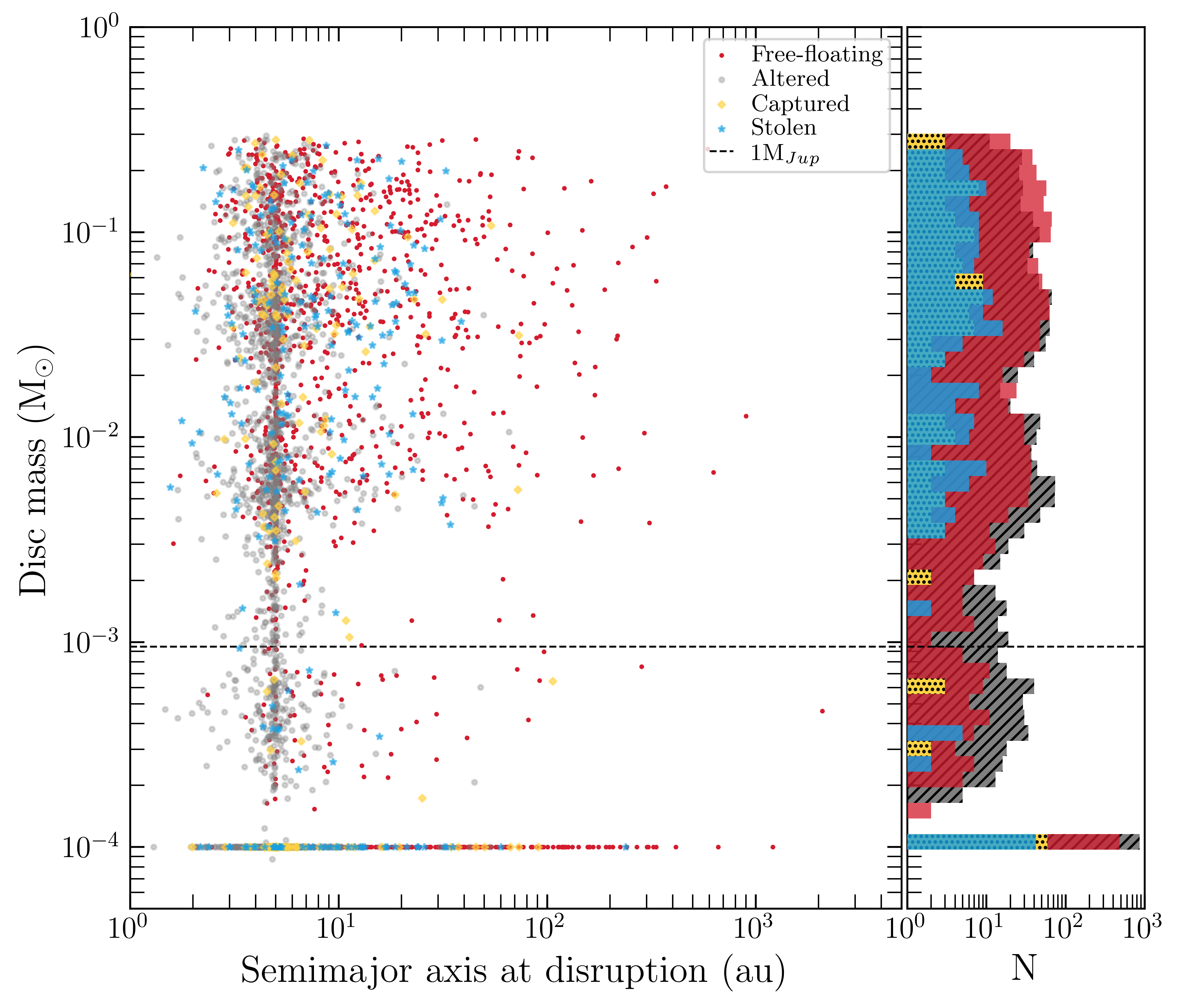}
    \caption{The mass of the hypothetical disc versus the semimajor axis of the planet when it is altered, captured or stolen, or when a planet becomes free-floating, in high density ($\tilde{\rho} = 10^4$\,M$_\odot$\,pc$^{-3}$) simulations. The disc mass is the total remaining mass in the disc, and the semimajor axis is the instant the planet undergoes an interaction. If all of the gas from the disc has been evaporated, the disc is assigned a mass of $10^{-4}$M$_\odot$ in the plot so that it can be visualised on the logarithmic axis.}
    \label{fig:sma_mass}
\end{figure}

In Fig.~\ref{fig:sma_mass} we show the mass of the protoplanetary disc versus the semimajor axis at the instant that the planet's orbit is disrupted. To visualise all the systems on this logarithmic scale plot, where all of the mass has been evaporated from the disc before disruption, we  assign the disc a mass of $10^{-4}$M$_\odot$ (almost no surviving discs have a mass this small).

\subsection{Moderate initial stellar density ($\tilde{\rho} = 10^2$\,M$_\odot$\,pc$^{-3}$)}

We now focus on the simulations with moderate initial stellar densities ($\tilde{\rho} = 10^2$\,M$_\odot$\,pc$^{-3}$). This simulations undergo the same subviral collapse and violent relaxation, but on a longer timescale. The erasure of substructure occurs within the first 1\,Myr (rather than 0.1\,Myr in the dense simulations) and the collapse to a bound star cluster occurs over 5\,Myr (rather than 0.5 --1\,Myr in the dense simulations).

\begin{figure}
    \centering
    \includegraphics[width=\columnwidth]{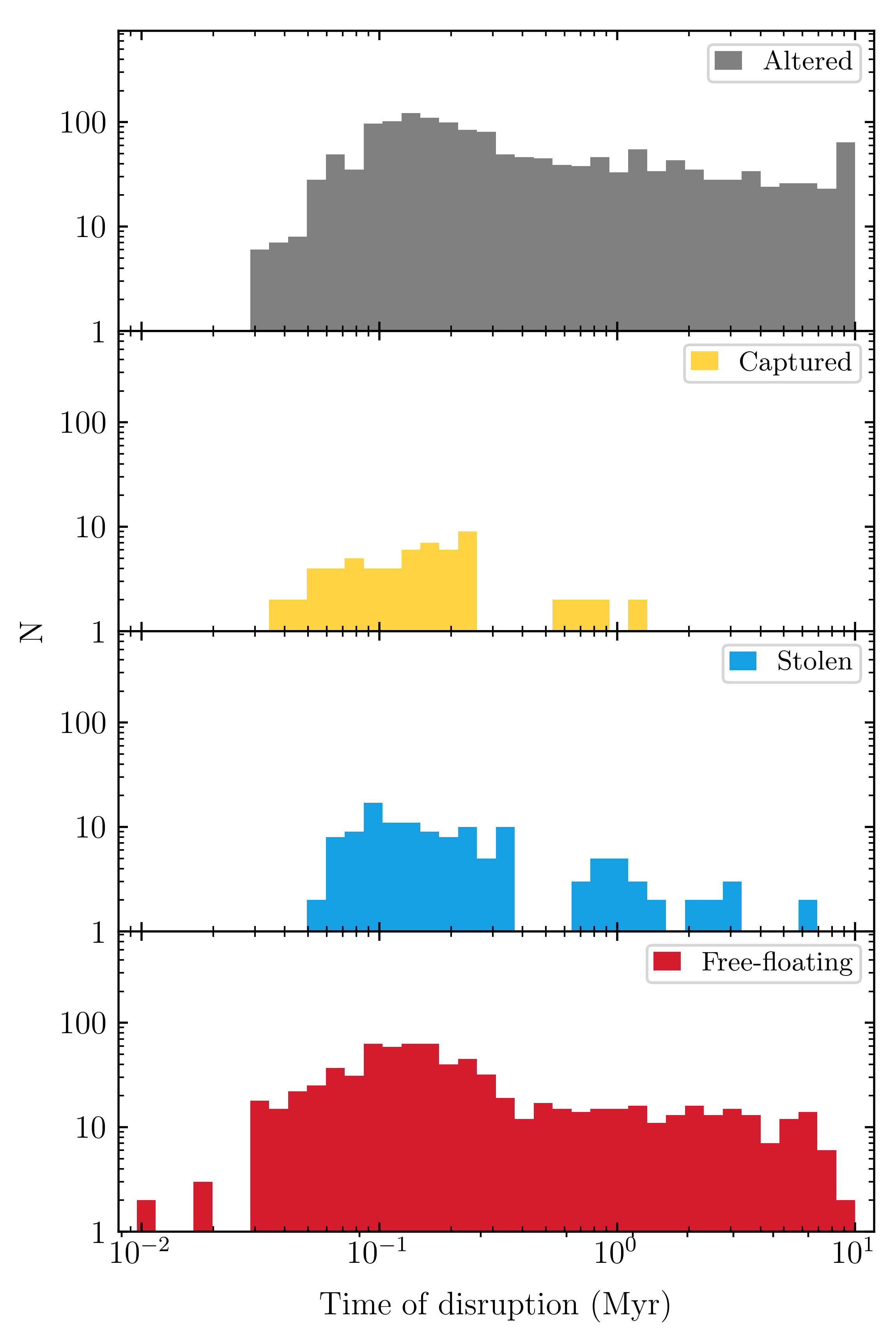}
    \caption{Histograms of the times at which planets are altered, captured, stolen or liberated in our simulations with moderate initial stellar densities ($\tilde{\rho} = 10^2$\,M$_\odot$\,pc$^{-3}$). The top panel shows planets that remain bound to their host star, but whose eccentricity has changed by more than 0.1, and/or semimajor axis has changed by more than 10\,per cent of the original value. The second panel shows planets that were captured, the third panel shows planets that were stolen by another star and the bottom panel shows planets that have become free-floating.}
    \label{fig:timescales_lowdensity}
\end{figure}

  In Fig.~\ref{fig:timescales_lowdensity} we plot the histograms of the times at which the planets are disrupted, and as we would expect this occurs to fewer systems overall (2515 out of 10\,000, compared to 4257 in the high-density simulations) and at later times in the simulation.

\begin{figure}
    \centering
    \includegraphics[width=\columnwidth]{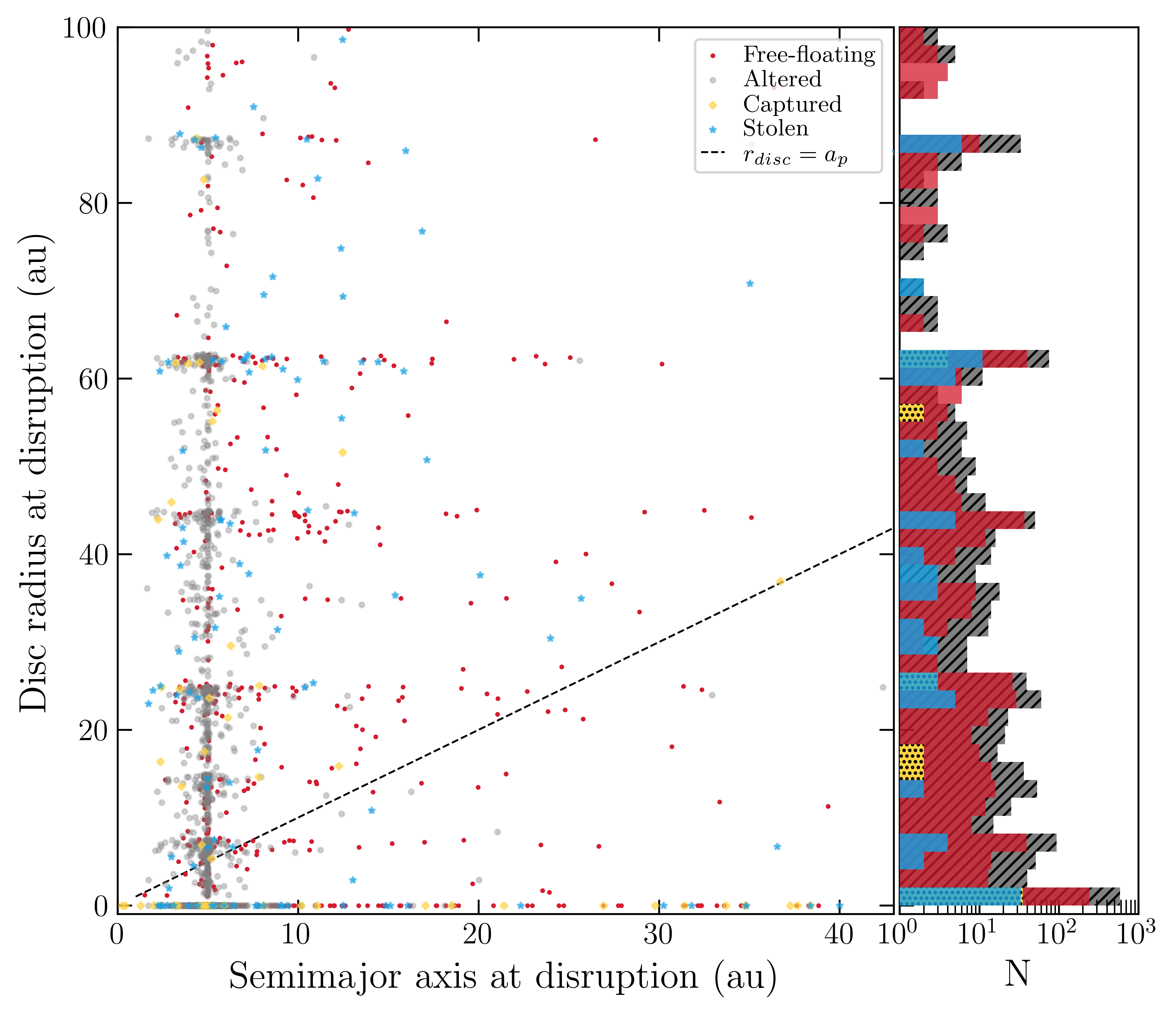}
    \caption{The radius of the hypothetical disc versus the semimajor axis of the planet when its orbit is altered, or when a planet is captured or stolen, or when a planet becomes free-floating, in our simulations with moderate initial stellar densities ($\tilde{\rho} = 10^2$\,M$_\odot$\,pc$^{-3}$). The disc radius is the outer radius of the disc, and the semimajor axis is the instant the planet undergoes an interaction. The dashed line shows $r_{\rm disc} = a_p$; where $a_p > r_{\rm disc}$, the gas in the disc has already been photoevaporated before the planet has undergone a significant dynamical encounter. The histogram indicates the numbers of planets in each category with a disc radius at the instant of disruption. }
    \label{fig:sma_rad_lowdensity}
\end{figure}

  In addition to there being fewer disruptive encounters that affect fully-formed planets, photoevaporation of the gas component of the protoplanetary discs is also less potent, with FUV radiation fields typically a factor of 100 lower in the lower-density simulations. In Fig.~\ref{fig:sma_rad_lowdensity} we show the disc radius at the time of planetary disruption versus the semimajor axis of the planet. We also show a histogram of the disc radii at the point the planets are disrupted. We find that of the 2515 disrupted planetary systems, 1043 (41\,per cent) have a disc radius lower than the planet's semimajor axis at the instant of disruption, and 883 (35\,per cent) have disc radii of zero similar to the high density simulations.  

\begin{figure}
    \centering
    \includegraphics[width=\columnwidth]{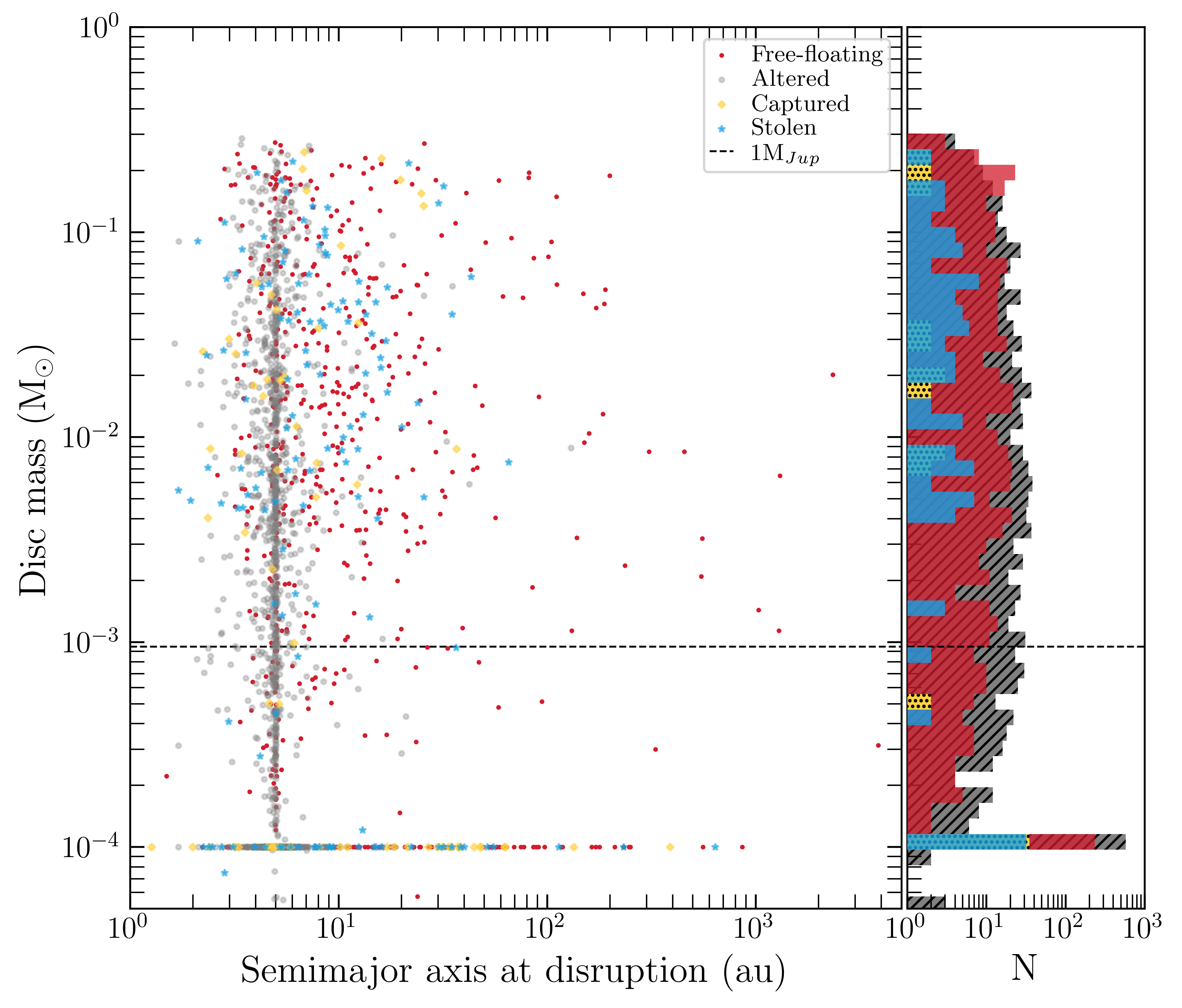}
    \caption{The mass of the hypothetical disc versus the semimajor axis of the planet when it is altered, captured or stolen, or when a planet becomes free-floating, in our simulations with moderate initial stellar densities ($\tilde{\rho} = 10^2$\,M$_\odot$\,pc$^{-3}$). The disc mass is the total remaining mass in the disc, and the semimajor axis is the instant the planet undergoes an interaction. If all of the gas from the disc has been evaporated, the disc is assigned a mass of $10^{-4}$M$_\odot$ in the plot so that it can be visualised on the logarithmic axis.}
    \label{fig:sma_mass_lowdensity}
\end{figure}

  In Fig.~\ref{fig:sma_mass_lowdensity}, we show the disc mass versus semimajor axis at the time of planetary disruption, and as in the case of the high density simulations, the majority of surviving discs contain enough gas to comfortably form Jupiter-mass planets.

Whilst there are fewer disruptive events in these lower density simulations, and fewer discs affected by photoevaporation, photoevaporation still dominates over disruption and occurs at even lower stellar densities than the regimes we model here \citep{Adams04,Adams06,Nicholson19a,Parker21a}.

\section{Discussion}

Unsurprisingly, our results are sensitive to the initial stellar density in the simulations. In the most dense star-forming regions, 40\,per cent of planets experience a disruptive encounter; either their orbit is significantly altered, or they are captured/stolen by another star, or they become free-floating. Of the disrupted planets, however, 44\,per cent of these systems have a disc that has a smaller semimajor axis than the planet when it is disrupted.

However, the proportion of disrupted planets that have a disc radius less than the planet's semimajor axis is similar (42\,per cent) in the moderate density simulations. The reason for this is that both dynamical encounters and photoevaporation have an inverse square dependence on the distance between stars. So, although fewer systems are affected overall in the lower-density simulations, we would expect significant gas-loss from discs in populous star-forming regions \emph{before} disruptive encounters with passing stars.

We emphasise that sub-mm dust particles are unlikely to be entrained in the wind launched by the incident radiation on the disc, and so significant amounts of solids will still be available for planet formation \citep{Haworth18b}. However, the decrease in the gas radius of the disc is likely to be followed by the dust radius \citep{Sellek20}, and so more material could be placed on smaller radii around their parent stars.

Previous studies have shown that whilst more than half of the discs lose all of their gas due to photoevaporation \citep{Scally02,Adams04,Nicholson19a,Parker21a}, Fig.~\ref{fig:sma_mass} demonstrates that those discs that do still contain gas usually have more than 1\,M$_{\rm Jup}$ of material (note the points above the horizontal dashed line). The presence of significant amounts of gas in these discs means that they could form gas giants, and there would also be significant amounts of dust with which to form terrestrial planets.

We suggest that the combination of gas mass-loss from protoplanetary discs that would otherwise form gas giant planets, followed by a sculpting of the disc and overconcentration of solids at smaller radii in discs could result in a very different population of planets than if systems formed without the influence of photoionising radiation. \citep[See also][who show that external photoevaporation can suppress accretion onto planets, as well as disrupting subsequent planetary migration.]{Winter22}

Many of the systems discovered by \emph{Kepler} are notable in that they contain systems of tightly packed super-Earth/mini-Neptune mass planets at very small semimajor axes \citep[e.g.][]{Mulders15}. Whilst there is considerable debate regarding the composition of these planets, it is clear that most of them are not gas giants like Jupiter and Saturn in our Solar system. Furthermore, it is currently unclear how so much material is assembled at such close radii to their host stars -- with some authors suggesting that the presence of unseen, distant giant planets could be responsible \citep{Hands16,Hansen17}.   

We speculate that some of these planetary systems could be the legacy of photoevaporation in the stellar birth environments of their host stars. If external photoevaporation drives the disc dust radii inwards \citep{Sellek20}, and this material is prevented from accreting onto the host star \citep[e.g. by internal FUV/XUV radiation from the parent star,][]{Picogna19}, then we might expect a significant build-up of solids in the inner au of the disc.

However, we also note that several recent studies have used RV measurements to demonstrate the presence of distant gas giant planets around some of the compact \emph{Kepler} systems \citep{Zhu18,Bryan19,Mills19,Zhu21,Chachan22,Smith22}. In these systems, perhaps the Jupiter-mass planets form quickly (e.g. through disc instabilities) before photoevaporation acts on the disc to form the super Earths, or these systems did not experience photoevaporation, and instead the outer giants aid and abet terrestrial planet formation in the inner regions of the disc.

\section{Conclusions}
\label{sec:conclusions}

We present $N$-body simulations of star-forming regions and determine when single Jovian-mass planets are stolen, captured, significantly altered or become free-floating. We then run a post-processing analysis on these simulations where we assume a protoplanetary disc is subject to photoevaporation from nearby massive stars, in order to determine the timescale for planetary disruption compared to disc destruction. Our conclusions are the following:

(i) Almost half (44\,per cent) of our stars that host $M_{\rm Jup}$ planets have smaller disc radii than the semimajor axis of the planet \emph{before} the planet orbit is disrupted. This means that planets that are disrupted in dense, populous star-forming regions would likely be ice giants or super Earths, rather than fully fledged gas giants.

(ii) The gas component in a significant number of discs ($\gtrsim$50\,per cent) is destroyed, but as almost no dust is lost to external photoevaproation \citep{Haworth18b} the discs retain enough mass for terrestrial planet formation ($>10^{-2}$M$_\odot$). However,  this mass is concentrated at smaller ($<10$\,au) radii, due to the evolution of the disc as it experiences photoevaporation \citep{Sellek20}. 

(iii) We speculate that the formation of some of the tightly-packed terrestrial planets discovered by \emph{Kepler} could have been aided and abetted by external photoevaporation from massive stars, which would cause the disc radii to move inwards, concentrating the solids. We might expect these planetary systems to be devoid of distant giant planets (which would not have been able to grow to Jovian masses due to a lack of gas). However, many of the \emph{Kepler} systems have recently been shown to harbour more distant Jovian-mass planets, which are presumably gas-rich.

(iv) If a significant majority of the \emph{Kepler} systems do harbour distant Jovian planets, this strongly argues against photoevaporation of protoplanetary discs being a dominant process in planet formation, which it turn would imply that most exoplanet host stars formed in low-mass (and/or very low-density) star-forming regions. 


\section*{Acknowledgements} 
 We thank the anaonymous referee for their comments and suggestions on the original manuscript. ECD was supported by the UK Science and Technology Facilities Council in the form of a
 PhD studentship. RJP acknowledges support from the Royal Society in the form of a Dorothy Hodgkin Fellowship. For the purpose of open access, the authors have applied a Creative Commons Attribution (CC BY) license to any Author Accepted manuscript version arising.


\section*{Data Availability}
The data underlying this article will be shared on reasonable request to the corresponding author.




\bibliographystyle{mnras}
\bibliography{general_ref} 




\label{lastpage}
\end{document}